\documentclass[12pt,journal, onecolumn, draftclsnofoot]{IEEEtran}
\include{psfig}
\usepackage{psfig}
\usepackage{psfrag}
\usepackage{amssymb}
\usepackage{graphicx}
\usepackage{cite}
\usepackage{subfigure}
\usepackage{tikz}
\usetikzlibrary{calc}
\usepackage[tbtags]{amsmath}
\usepackage{epsfig}
\usepackage{stfloats}
\usepackage{array}
\usepackage{color}
\usepackage{pgfplots}
\pgfplotsset{compat=newest}
\pgfplotsset{plot coordinates/math parser=false}

\usetikzlibrary{plotmarks}

\newcommand{\var}{\sigma^2}

\newcommand{\E}{\mathrm{E}}

\newcommand{\bqn}{\begin{equation}}
\newcommand{\eqn}{\end{equation}}
\newcommand{\bqa}{\begin{eqnarray}}
\newcommand{\eqa}{\end{eqnarray}}
\newcommand{\bqas}{\begin{eqnarray*}}
\newcommand{\eqas}{\end{eqnarray*}}

\newcommand{\av}{{\bf a}}

\newcommand{\uv}{{\bf u}}
\newcommand{\wv}{{\bf w}}

\DeclareGraphicsExtensions{.pdf,.png,.jpg}

\begin{document}
\title{Simple Semi-Distributed Lifetime Maximizing Strategy via 
Power Allocation in Collaborative Beamforming for Wireless 
Sensor Networks}
\author{Mohammed F. A. Ahmed, Sergiy A.~Vorobyov
\thanks{Mohammed F. A. Ahmed is with the Computer, 
Electrical, and Mathematical Science and Engineering (CEMSE) Division, 
King Abdullah University of Science and Technology (KAUST), Thuwal, 
Makkah Province, Kingdom of Saudi Arabia.
Sergiy A. Vorobyov is with {the Dept. of Signal Processing and Acoustics, 
Aalto University. He is on leave from} the Department of Electrical and 
Computer Engineering, University of Alberta, 9107-116 St.,
Edmonton, Alberta, T6G~2V4 Canada. Emails: {\tt
m.ahmed@kaust.edu.sa} and {\tt svor@ieee.org}. }} 

\maketitle
 
\begin{abstract}
Energy-efficient communication is an important issue in wireless
sensor networks (WSNs) consisting of large number of energy-constrained sensor
nodes. Indeed, sensor nodes have different energy budgets assigned
to data transmission at individual nodes. Therefore, without
energy-aware transmission schemes, energy can deplete from sensor
nodes with smaller energy budget faster than from the rest of the
sensor nodes in WSNs. This reduces the coverage area as well as the
lifetime of WSNs. Collaborative beamforming (CB) has been proposed
originally to achieve directional gain, however, it also
inherently distributes the corresponding energy consumption over
the collaborative sensor nodes. In fact, CB can be seen as a
physical layer solution (versus the media access control/network
layer solution) to balance the lifetimes of individual sensor
nodes and extend the lifetime of the whole WSN. However, the
introduction of energy-aware CB schemes is critical for {extend}ing
the WSNs lifetime.

In this paper, CB with power allocation (CB-PA) is developed to
{extend} the lifetime of a cluster of collaborative sensor nodes by
balancing the individual sensor node lifetimes. A novel strategy
is proposed to utilize the residual energy information available
at each sensor node. It adjusts the energy consumption rate at
each sensor node while achieving the required average
signal-to-noise ratio (SNR) at the destination. It is a
semi-distributed strategy and it maintains average SNR. Different 
factors affecting the energy consumption are studied as well. 
Simulation results show that CB-PA outperforms CB with Equal Power 
Allocation (CB-EPA) in terms of {extend}ing the lifetime of a 
cluster of collaborative nodes.
\end{abstract}


\section{Introduction}
Typically, sensor nodes in wireless sensor networks (WSNs) are
equipped with batteries or energy harvesting devices as energy
sources \cite{Paradiso2005}-\cite{Penella2009}. Since batteries
have limited capacity and harvesting devices provide very small
amount of energy, energy is a scarce resource at sensor nodes
\cite{Alippi2009}, \cite{Simjee2008}. As a result, practical WSN
implementations require energy efficient schemes to conserve
energy and extend the network lifetime {{as much as possible}}. 
Another challenge when
addressing energy conservation in WSNs is the unequal energy
consumption rate for individual sensor nodes. Sensor nodes 
are typically assigned different sensing, data processing, and
communication tasks depending on the deployment scenario
\cite{Sinha2001}, \cite{Park2005}. Therefore, even if sensor nodes
start with the same energy budget once the WSN deployed, the
surrounding environment determines the priorities and the loads of
different tasks assigned for different sensor node, which causes
that different sensor nodes deplete energy at
different rates \cite{Chen2009}, \cite{PRVor}. Therefore, sensor
nodes may have different residual energies at their batteries with
progress of WSN operation time.

Energy-efficient communication can be partially addressed in the
physical (PHY) layer by implementing collaborative beamforming
(CB) \cite{Ochiai2005}-\cite{Ahmed2010}, where a single-hop
transmission is established when a number of sensor nodes in a
cluster adjust their carrier phases to cancel out the phase
differences due to propagation delays and, thus, signals add
coherently at the base station/access point (BS/AP). It is an
efficient communication technique in situations when distributed
low-power source nodes are required to achieve long distance
transmissions toward far away BS/AP. Moreover, it is a scalable
technique that can be applied to very large scale WSNs
effectively. The effect of the scalability on CB performance is,
however, totally different from the case of multi-hop transmission
\cite{Le2007}, \cite{Madan2006}. Indeed, in the case of multi-hop
transmission, the increase of the number of sensor nodes
introduces more problems to the design \cite{Alazzawi2008},
\cite{Abdulla2011}. In the CB case, however, the increase of the
number of sensor nodes improves the CB performance. Specifically,
it enables the sample beampattern of CB to approach the average
beampattern \cite{Ochiai2005}, \cite{Ahmed2009} as well as it
helps to spread the energy consumptions over more sensor nodes in
WSN. Thus, the important difference between the CB technique and
the multi-hop transmission in terms of energy-efficient
communication is that the CB technique inherently spreads the
energy cost over a group of collaborative sensor nodes
\cite{Tokunaga2008}.

Two requirements must be satisfied while implementing the CB,
namely, phase synchronization and information sharing. The first
requirement implies that the carrier signals from different sensor
nodes must be synchronized in phase and frequency to achieve
coherent transmission. Distributed schemes for estimating the
initial phases of the local node oscillators in WSNs have been
introduced in \cite{Mudumbai2006}, \cite{Brown2008} . A
synchronization algorithm developed in \cite{Mudumbai2006} uses a
simple 1-bit feedback iterations, while other methods developed in
\cite{Brown2008} and \cite{QWang2008} are based on the
time-slotted round-trip carrier synchronization approach. The
second requirement implies that information to be transmitted must
be available at all sensor nodes before the CB transmission step
for enabling simultaneous transmission of the same data symbol.
Information sharing can be achieved in a straightforward manner
via broadcasting a common symbol from one sensor node to other
collaborative sensor nodes if the WSN has only one source node,
and over orthogonal channels in WSNs with multiple source nodes
\cite{Dong2009}.

The CB with equal powers transmitted by sensor nodes, which will be
called hereafter CB with Equal Power Allocation (CB-EPA), can still
lead to the situation where sensor nodes with smaller energy
budgets drain out of energy faster than the rest. Energy
consumption is considered in some particular CB schemes existing
in the literature. Specifically, the power-saving effect of the CB
technique is studied in \cite{Tokunaga2008} with the assumption of
free-space channels between the sensor nodes and the targeted
BS/AP. In \cite{Feng2010}, sensor node scheduling is proposed for
CB where the participating sensor nodes are selected in each round
of transmission to balance the remaining energies at all sensor
nodes. A game-theoretic approach is used in \cite{Betz2008} for power
control among different WSN clusters utilizing CB for
transmission. The corresponding algorithms are addressing
different aspects affecting the WSN lifetime such as scheduling
and power allocation among clusters of sensor nodes. However, for
the corresponding algorithms, information about all sensor
nodes in a cluster is required at a central point to, for example,
schedule the sensor nodes appropriately in order to achieve a
balanced remaining energy across the cluster. Distributed
algorithms, which allow to avoid an overhead in WSNs, are however
of significant importance.

In this paper\footnote{Some initial results have been reported in
\cite{Asilomar11}.}, the CB technique with power allocation (CB-PA) 
is developed to {extend} the lifetime of a cluster of collaborative
sensor nodes in a WSN. A novel strategy is proposed to achieve this goal by
balancing the lifetime of individual sensor nodes instead of
balancing their energy consumption. Residual energy information
(REI) available at each sensor node is utilized to adjust the
transmission power of each sensor node while achieving the
required average signal-to-noise ratio (SNR) at the targeted
BS/AP. The proposed algorithm requires significantly lower
overhead and computational complexity as compared to the
centralized algorithms. Note that in WSNs, distributed algorithms
are generally preferred over centralized protocols even when they
offer only a sub-optimal solution \cite{Cohen2010}. Simulation
results show that CB-PA outperforms CB-EPA in terms of the 
lifetime of a cluster of sensor nodes and the SNR improvement at
the BS/AP. Different factors affecting the performance of the
proposed power allocation strategy are investigated. For example,
we show that the initial energy distribution across sensor nodes
affects the lifetime of WSN. We also show that implementing the
multi-link CB \cite{Ahmed2010} is preferable in terms of the
energy-efficiency than the single-link CB \cite{Ochiai2005},
\cite{Ahmed2009}. Moreover, reduction of the required bit rate
[bits/sec/Hz] leads to a considerable increase in the WSN
lifetime.

The rest of the paper is organized as follows. The WSN geometric 
and signal models are introduced in the next section. Section III 
motivates and introduces an energy
consumption model for WSNs. The notion of the lifetime of a cluster
of WSN is also introduced. A novel partially distributed and
efficient power allocation strategy for CB in WSN is introduced
and investigated in Section IV. Section V further discusses the
effect of different factors such as, for example, initial energy
distribution in a cluster of WSN and consideration of multi-link
CB versus single-link CB on the WSN lifetime. The latter studies
are also performed in terms of simulations in Section~VI.
The paper ends with conclusions. This paper is reproducible
research \cite{Rep} and the software needed to generate the
simulation results will be provided together with the paper upon
its acceptance.

\section{Geometric and Signal Models}
\subsection{Geometric Model}
\begin{figure}[t]
\centering
\includegraphics{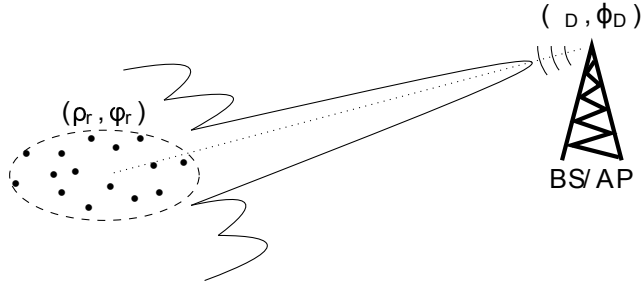}
\caption{Cluster of collaborative sensor nodes transmitting to a
BS/AP using CB.}
\label{Fig:Model}
\end{figure}

Fig.~\ref{Fig:Model} shows the geometric model of a cluster of $N$
sensor nodes $c_{i}, i = 1,2, \ldots, N$, which are randomly
placed over a  plane.\footnote{We will use $i$ and $p$ to denote
the sensor index and later in the case of multi-link CB we will
use $k$ and $l$ to denote the BS/AP index.} Sensor nodes serve as
collaborative nodes to transmit data to a remote BS/AP, denoted as
${\cal D}$. We use polar coordinates $(\rho,\phi)$, where $\rho$
is the distance from the origin to a given point and the angle
$\phi$ represents the azimuth direction. Let us assume that the
destination BS/AP is located at $(\varrho_D,\varphi_D)$ and is in
the same $x$-$y$ plane in which the sensor nodes are located. Let
the $r$th sensor node be located at $\left( \rho_i , \phi_i
\right)$. The polar coordinates of sensor nodes are chosen
independently and randomly according to an arbitrary spatial
distribution. Each sensor node is aware of its location and is
able to communicate with other sensor nodes in the cluster using
links with low power consumption for information sharing and
synchronization. We also assume that sensor nodes are frequency 
synchronized and frequency drift effects are negligible \cite{Brown2008}.

\subsection{Signal Model}
Collaborative sensor nodes are assumed to have access to the
data/measurements to be transmitted to the intended BS/AP.
Information sharing is performed by broadcasting the data from one
source node to all other nodes in its coverage area
\cite{Ochiai2005}, \cite{Ahmed2010}, \cite{Dong2009}. In the case
of multiple source nodes, information sharing can be achieved over
orthogonal channels in frequency, time, or code to avoid
collisions. Note that the use of time orthogonal channels requires
scheduling which can be achieved by an appropriate MAC protocol or
using the BSs/APs as a central scheduler. Alternatively, the
existing collision resolution schemes can be used \cite{Lin2005}.
We assume that the power used for broadcasting the data by the
source node is high enough so that each collaborative node $c_i$
can successfully decode the received symbols from the source node.

During the CB step, collaborative sensor nodes use the same
codebook with zero mean, unit power, and independent symbols,
i.e., $\E \left\{ z_n \right\} = 0$, $\left| z_n \right|^2 = 1$,
and $\E\left\{z_n \cdot z_m \right\} = 0$ for $n \ne m$ where $z_n$
stands for the $n$th symbol from the codebook and $\E \{ \cdot \}$
denotes expectation.

Considering a discrete-time system, each collaborative node
$c_{i}$,  $i = 1,2, \ldots, N$ from a cluster of $N$ nodes in WSN
transmits the signal
\begin{equation}
t_i  =  z \cdot w_i \cdot  e^{j \psi_i}, \quad   i = 1,2, \ldots, N
\end{equation}
where $z$ is the transmitted symbol, $w_{i}$ is the CB real valued
weight for senor node $c_{i}$. The real-valued $w_{i}$ controls
the transmission power of sensor node $c_{i}$ since $P_{i} =
w_{i}^2$, and $\psi_i$ is the initial phase of the carrier. The
initial phases of the carriers are adjusted using the well known
algorithms \cite{Mudumbai2006}-\cite{Brown2008}, while we are
interested here in finding the real-valued $w_{i}$, i.e., in
controlling the transmit power. However, it is worth noting that
although we refer to $w_{i}$ as the CB weight for simplicity, the
actual CB weight at the collaborative node $c_i$ is a
complex-valued beamforming weight, that is, $w_{i} \cdot e^{j
\psi_i}$.

Under the assumption of far-field region, i.e., ${\varrho_D} \gg
\rho_i$, the Euclidean distance between the collaborative node $c_i$
and a point $(\varrho_D,\phi)$ in the same plane is given as
\begin{equation}
\delta_i (\phi) \approx \varrho_D - \rho_i \cos(\phi-\phi_i).
\end{equation}
and the corresponding phase shift is
\begin{equation}
\theta_i ( \phi) = \frac{2\pi}{ \lambda} \cdot \delta_i (\phi).
\end{equation}
Thus, using the knowledge of the node location and the direction
of the destination ${\cal D}$, the initial phase of each sensor
node carrier must be set as (see the closed-loop scenario in
\cite{Ochiai2005})
\begin{equation}
\psi_i = -  \frac{2\pi}{ \lambda} \cdot \delta_i
(\varphi_{\cal D})
\end{equation}
in order to achieve coherent combining at the destination ${\cal
D}$. In (4), $\lambda$ is the wavelength. Alternatively,
synchronization can be performed without any knowledge of the node
locations using time-slotted round-trip synchronization
\cite{Brown2008} or feedback-based synchronization
\cite{Mudumbai2006}. Sufficient level of phase accuracy and
reliability is achieved with these techniques at the cost of small
overhead. Then, the received signal at angle $\phi$ from all
collaborative nodes in a cluster can be written as
\begin{eqnarray}
g (\phi) = z \cdot \sum\limits_{i=1}^{N} w_{i} \cdot e^{j \psi_i}
\cdot a_{i} \cdot e^{ j \theta_i ( \phi)}  + n
\end{eqnarray}
where {$a_i$ is the channel coefficient between the $i$th collaborative 
node and the destination ${\cal D}$ and }$n \sim {\cal CN} ( 0, 
\sigma_{n}^2 )$ is the AWGN at the direction $\phi$ with variance 
$\sigma_{n}^2$. Practical implementation for CB in WSNs requires the 
sensor nodes to be deployed over a large cluster area to achieve CB 
beampattern with narrow mainlobe \cite{Ochiai2005}. Consequently, 
the distances between sensor nodes are assumed to be very large 
relative to the carrier wavelength at the operating frequency and the 
channel coefficients are uncorrelated. The received signal at the 
BS/AP, ($\phi = \varphi_{\cal D}$), can be then expressed as
\begin{eqnarray}
y = z \cdot \sum\limits_{i=1}^{N} w_{i} \cdot a_{i} + n = z \ \wv^T 
\av + n.
\label{Eq:y1Ch5}
\end{eqnarray}
\noindent where $\wv = \left[ w_1, \ w_2, \ \dots  \ , w_N \right]^{T}$ 
is the weights vector, $w_i \in [0,w_{\rm max}]$, $\av = \left[ a_1, \ 
a_2, \ \dots  \ , a_N \right]^{T}$ is the channel vector, and $(\cdot)^T$ 
denotes the transpose of a vector. The corresponding SNR $\gamma$ 
is given by
\begin{eqnarray}
\gamma  = \frac{|\av^T \wv|^2}{\var_n}.
\end{eqnarray}

\section{Energy Consumption Model in a Cluster of WSN}
In this section, we develop an energy consumption model for a 
cluster of WSNs with focus on the energy dissipation corresponding 
to CB {weights}. We also introduce specific definition for the
lifetime of a cluster of collaborative sensor nodes implementing
CB transmission.

\subsection{Energy Consumption Model}
A popular energy consumption model has been introduced in the
literature for transmission in WSNs \cite{Feng2010},
\cite{Perillo2009}, \cite{Mallinson2008}. This model is commonly
used for the analysis of the media access control and
network layer protocols designed in order to maximize the network
lifetime. The overall energy consumption consists of the energy 
consumed by the radio-frequency (RF) transceiver hardware in both 
transmission and reception phases. The consumed energy in the 
transmitter is dissipated in the circuit electronics and the power amplifier. 
On the other hand, the receiver does not have a power amplifier and
the energy is dissipated in the circuit electronics only.

The energy consumed by the transmitter can be expressed as
\begin{eqnarray}
 E_{\rm Tx} = E_{e,\rm Tx} + E_{\rm a}
\end{eqnarray}
where  $E_{e,\rm Tx}$ represents the energy consumption of
the transmitter electronics and $E_{\rm a}$ is the energy consumed
by the transmit power amplifier. The energy $E_{e,\rm Tx}$
is consumed in the transmitter hardware, including the oscillator,
frequency synthesizer, mixers, filters, baseband processor, etc.
This energy is considered to be constant for a specific hardware. 
The energy consumed in the power amplifier of $c_i$'s collaborative 
sensor node depends on the CB weight $w_i$ assigned to this sensor 
node. The transmitted power from each individual sensor node is 
$P_{i} = w_{i}^{2}$. A time-slotted transmission is considered and
sensor nodes transmit data to the targeted BS/AP over time slot
$t$ of length $T$ seconds. The energy $E_{\rm a}$ consumed during 
one time-slot is then given by
\begin{eqnarray}
E_{\rm a} = e_{i}(t) =  w_{i}^{2} \cdot T = P_{i} \cdot T.
\end{eqnarray}

The total transmitted power $P^{\rm Tx}= \sum_{i=1}^{N} P_{i}$ has 
to compensate the attenuation due to propagation distance. If 
free-space propagation is considered, $P^{\rm Tx}$ can be written 
(in dB) as
\begin{eqnarray}
P_{\rm Tx} = P_{\rm Rx} + {PL}_0 + 10 \cdot \alpha \cdot 
\log_{10} \left( \frac{d}{d_0} \right) 
\end{eqnarray}

\noindent where $P_{\rm Rx}$ [dB] is the received power at the 
destination corresponding to the received SNR $\gamma$, $d$ is the 
distance over which data is being communicated, ${\rm PL}_0$ [dB] is 
the path-loss at the nominal distance $d_0$, and $\alpha$ is the 
path loss exponent. 

The energy consumed at the receiver can be expressed as
\begin{eqnarray}
E_{\rm Rx} = E_{e,\rm Rx}
\end{eqnarray}
where $E_{e,\rm Rx}$ represents the energy consumption
of the receiver electronics. Similar to $E_{e,\rm Tx}$,
$E_{e,\rm Rx}$ is considered constant for specific
hardware. In the following analysis, we only consider the energy 
consumption corresponding to the CB transmission in WSNs 
and, thus, we neglect $E_{e,\rm Tx}$ and $E_{e,\rm Rx}$ {{ because 
it is the same for both CB-EPA and CB-PA}}.

\subsection{Lifetime of a Cluster of Sensor Nodes}
The network lifetime has many definitions in the literature.
However, the common understanding of the network lifetime is the
time period until the WSN stops performing its assigned tasks
\cite{Noori2006}. Particular definitions depend on the criteria
according to which the network is recognized as not functional.

In the context of WSNs, where a cluster of collaborative nodes
utilizes CB for transmission, two main tasks should be performed
in order for the network to be functional. First, sensor nodes are
required to collect information from the surrounding environment.
Second, sensor nodes are required to communicate the collected
information to the BS/AP using CB. Thus, we define the lifetime of
a cluster of collaborative sensor nodes, denoted hereafter as
$\tau$, as the time period during which the number of alive sensor 
nodes is larger than a certain value and the collaborative sensor
nodes are able to achieve acceptable SNR at the targeted BS/AP.

\section{Power Allocation Strategy for CB}
In this section, we propose a low-complexity and semi-distributed
power allocation scheme for CB in a cluster of collaborative nodes
of WSN aiming at balancing the lifetimes of individual sensor
nodes in order to maximize the lifetime of the cluster.

\subsection{Power Allocation Strategy}
Proper operation of WSN requires that the achieved SNR $\gamma$ at 
the BS/AP does not fall below a certain level for correct reception 
of the signal. Power allocation that maximizes the beampattern 
gain at the BS/AP under the total and individual power constraints
can be used to design the CB weights. The corresponding optimization 
problem can be stated as
\begin{eqnarray}
\begin{aligned}
& \underset{\wv}{\text{maximize}} & & |\av^T \wv|^2  \\
& \text{subject to} & & \sum\limits_{i=1}^{N} w^2_i = P_{\rm tot}   
\\
&  & & w^2_i \le  P_{\rm max} , \quad i=1, \dots, N
\end{aligned}
\end{eqnarray}
\noindent where $P_{\rm tot}$ denotes the total power to be transmitted 
from the sensor nodes at each round and $P_{\rm max}$ is the 
maximum transmitted power for each sensor node. Alternatively, the 
power allocation problem can be reformulated as extension of the 
lifetime of the network by minimizing the norm of the CB weights vector 
while insuring that the SNR $\gamma$ is larger than a predetermined 
average value $\bar{\gamma} \triangleq \E\{ \gamma \}$ at the intended 
BS/AP under the individual power constraints. The corresponding 
optimization problem can be written as
\begin{eqnarray}
\begin{aligned}
& \underset{\wv}{\text{minimize}} & & \| \wv \|^2  \\
& \text{subject to} & & \gamma  \ge  \bar{\gamma}   \\
&  & & w^2_i \le  P_{\rm max} , \quad i=1, \dots, N.
\end{aligned}
\end{eqnarray}

Despite the fact that the aforementioned optimization problems 
can be solved computationally very efficiently, {{to}} address them 
practically, it is required to find the optimum solution centrally at 
the BS/AP, which will be subsequently transmitted to the sensor 
nodes (each individual weight to the corresponding sensor node). 
This scenario requires very {{large}} transmission overhead 
between the sensor nodes and the BS/AP and thus is not suitable 
for the WSN applications because of the high communication 
complexity. Therefore, it is of great interest to solve 
the power allocation problem in a distributed fashion. 

Power allocation for CB should achieve the following purposes:
\begin{itemize}
\item CB weights should balance the lifetime of individual sensor
nodes instead of equalizing the energy consumed for individual
CB transmissions;
\item It should be guaranteed that the received SNR $\gamma$
at the targeted BS/AP achieves the predetermined average value
$\bar{\gamma}$ over the lifetime.
\end{itemize}

\noindent To achieve the aforementioned requirements, power 
allocation can be performed in two steps. Namely, the first step 
is to calculate a normalized CB weights based on the REI at 
each sensor node to balance the lifetime of individual sensor 
nodes. The second step is to find a scaling factor, i.e., maximum 
CB weight, to achieve the required average SNR $\bar{\gamma}$ 
at the targeted BS/AP.

Let the vector ${\boldsymbol u} = [u_1,u_2,$ $\ldots,u_N]^{T}, 
u_i \in [0,1],$ stands for the normalized CB weight vector and 
$w_{\rm max}$ is the maximum CB weight. Then, the CB weight 
vector can be found as
\begin{equation} \label{wrDef}
\wv = w_{\max} \cdot \uv,
\end{equation}
that is, the weight $w_i$ is a fraction of the maximum CB weight
given the corresponding normalized CB weight $u_i$. Also, let us
introduce the vector ${\boldsymbol e} = [e_1, e_2, \ldots,
e_N]^{T}$ as the REI vector, where $e_i \in [0,E_{\max}]$ and $E_{\rm
max}$ is the battery capacity, which is assumed to be initially
the same for all collaborative sensor nodes in a cluster.

In order to find the normalized CB weights, i.e., the vector
${\boldsymbol u}$, we assume that each sensor node can measure its
own REI $e_{i}$. The normalized CB weight vector can be then
designed to balance the lifetimes of different sensor nodes so
that larger CB weights are assigned to sensor nodes with larger
REIs. A simple and fully distributed way of obtaining the
normalized CB weights can be expressed as
\begin{eqnarray} \label{uDef}
u_i = \frac{ e_i }{ E_{\max} } , \quad i = 1,2, \dots, N.
\end{eqnarray}
We call such scheme distributed because the normalized weight at
one collaborative sensor node is independent of any information
required to compute the weight at another collaborative sensor
nodes in a cluster of WSN.

The scaling factor $w_{\max}$ corresponding to ${\boldsymbol
u}$ is used to adjust the average received SNR at the targeted
BS/AP and has to be also computed. The average SNR $\bar{\gamma}$
at the targeted BS/AP ${\cal D}$ can be found as
\begin{eqnarray} \label{gammaDef}
\bar{\gamma} = \E\{ \gamma \} = \frac{\E \left\{ \left|z \cdot
\sum\limits_{i=1}^N w_{i} \cdot a_{i} \right|^2\right\} }{
\sigma_{n}^2} .
\end{eqnarray}
Substituting \eqref{wrDef} into \eqref{gammaDef}, the average SNR
$\bar{\gamma}$ can be expressed after some straightforward
computations as a function of $w_{\max}$, that is,
\begin{eqnarray}
\bar{\gamma} \!\!\!&=&\!\!\!  \frac{ w_{\max}^2 }{\sigma_{n}^2}
\cdot \sum\limits_{i=1}^N  \E\left\{u_{i}^2 \right\} \cdot  \E
\left\{a_{i}^2 \right\}  \nonumber \\
&+&  \frac{ w_{\max}^2 }{\sigma_{n}^2} \cdot \sum\limits_{i=1}^N 
\sum\limits_{\substack{p=1, \\ p \ne r}}^N  \E\left\{u_{i} \right\}  
\cdot  \E \left\{a_{i} \right\}
 \cdot  \E\left\{u_p \right\}
 \cdot  \E\left\{a_p \right\}  \nonumber \\
\!\!\!&=&\!\!\!  \frac{ w_{\max}^2 }{\sigma_{n}^2} \cdot  \left( 
N \cdot  (\sigma_{\rm u}^2 + m_{\rm u}^2) \cdot  (\sigma_{\rm  a}^2 
+ m_{\rm a}^2) \right)  \nonumber \\
&+& \frac{ w_{\max}^2 }{\sigma_{n}^2} \left( N
\cdot (N-1) \cdot  m_{\rm u}^2  \cdot m_{\rm a}^2 \right)
\label{Eq:aveSNRCh5}
\end{eqnarray}
\noindent where $m_{\rm u}$, $\sigma_{\rm u}^2$, $m_{\rm a}$, and
$\sigma_{\rm a}^2$ are the mean and variance of the normalized CB
weights and the mean and variance of the corresponding channel
gains, respectively. Then, the maximum transmission CB weight
corresponding to the average SNR (\ref{Eq:aveSNRCh5}) at the
targeted BS/AP ${\cal D}$ can be found by solving
\eqref{Eq:aveSNRCh5} for $w_{\max}$, and it can be expressed as
\begin{equation}
\begin{split}
&w_{\max} =  \\
&\sqrt{ \frac{ \bar{\gamma} \cdot \sigma_{n}^2}{ N
\cdot (\sigma_{\rm u}^2 \cdot \sigma_{\rm  a}^2 + \sigma_{\rm a}^2
\cdot  m_{\rm u}^2  + \sigma_{\rm  u}^2  \cdot m_{\rm a}^2 ) + N^2
\cdot m_{\rm u}^2 \cdot m_{\rm a}^2 } }.
\end{split}
\end{equation}
The mean and variance of the normalized CB weights can be found
from the mean $m_{\rm e}$ and variance $\sigma_{\rm e}^{2}$ of the
REI as
\begin{eqnarray}
m_{\rm u} \!\!\!&=&\!\!\! \frac{  m_{\rm e} }{E_{\max}} \\
\sigma_{\rm u}^2 \!\!\!&=&\!\!\!  \frac{\sigma_{\rm e}^{2}
}{E_{\max}^2}
\end{eqnarray}
where $m_{\rm e}$ and $\sigma_{\rm e}^{2}$ can be obtained through
consensus or distributed estimation algorithms
\cite{Xiao2007}-\cite{Cattivelli2010}. Thus, also $m_{\rm e}$ and
$\sigma_{\rm e}^{2}$ are computed in a distributive manner, the
scaling factor $w_{\max}$ has to be communicated to all
collaborative sensor nodes in a cluster. Therefore, we call the
proposed power allocation technique semi-distributed.

The power allocation can be performed every time slot. It results
in an undesirable overhead since the normalization factor
$w_{\max}$ has to be computed and collected by individual CB nodes
at every time slot. However, the remaining energy at different
nodes is not changed dynamically for consecutive time slots. 
Thus, power allocation does
not need to be performed every time slot and the BS/AP can
instruct the sensor nodes to perform power allocation only when
needed to minimize the overhead.

An alternative way that allows to avoid calculating $w_{\max}$
is the following. Start the CB transmission with predetermined
value $w_{\max} = \hat{w}_{\max}$. The targeted BS/AP
instructs then the collaborative sensor nodes in a cluster to
increment/decrement $w_{\max}$ in predetermined steps to reach
an acceptable SNR at the BS/AP. Note that the number of sensor
nodes $N$ in the cluster should be large enough to achieve the
required SNR with  $w_{\max} \le \sqrt{P_{\rm max}}$ given by 
the specifications of the sensor node transmitter.

\subsection{Residual and Wasted Energies}
In addition to the lifetime notion, there are other important
notions such as the residual and wasted energies. The wasted
energy is naturally the energy that is left in sensor nodes in a
cluster of a WSN after the cluster is no longer able to perform a
prescribed function with a required quality of service, such as
SNR requirement at the targeted BS/AP. The wasted energy is as
well an important characteristic of WSN especially from the energy
efficiency point of view.

Let $e_{i}(0)$ be the initial energy budget dedicated to CB at
each sensor node $c_{i}$, $i = 1,2, \ldots, N$. Then the residual
energy at the sensor node $c_{i}$ at the end of the $t$-th
transmission round can be expressed as
\begin{eqnarray}
e_{i}(t) = e_{i}(0) - \sum\limits_{l = 1}^{t} e_{i} (l) .
\end{eqnarray}
The residual energy at the sensor node $c_{i}$ when the cluster of
sensor nodes dies, denoted as $\epsilon_{i}$, can be
correspondingly given as
\begin{eqnarray}
\epsilon_{i} = e_{i}(\tau) = e_{i}(0) - \sum\limits_{l = 1}^{\tau}
e_{i}(l)
\end{eqnarray}
that is the difference between the initial energy budget and the
energy used during all transmission round until the last round
$\tau$ after which the cluster of WSN dies.

Finally, the wasted energy at a whole cluster of sensor nodes
is defined as the total unused energy in the cluster when it dies
and it is given by
\begin{eqnarray}
\epsilon = \sum\limits_{i=1}^{N} \epsilon_{i}.
\end{eqnarray}
Then the wasted energy can be expressed as a percentage of the
total initial energy as
\begin{equation}
\epsilon_{\%} = \frac{\epsilon}{N \cdot m_{\rm e}} \times 100 \%.
\end{equation}

\subsection{CB with Equal Power Allocation}
The CB-PA needs to be compared to the CB-EPA, i.e., the CB for
which the CB weights are equal, that is, $u_i = 1$. The average
SNR $\bar{\gamma}$ for in the case of equal power CB can be
expressed as
\begin{eqnarray}
\bar{\gamma} \!\!\!&=&\!\!\! \E\{ \gamma \} = \frac{\E \left\{
\left|z \cdot \sum\limits_{i=1}^N w_{i} \cdot a_{i} \right|^2
\right\} }{\sigma_{n}^2} \nonumber \\
\!\!\!&=&\!\!\! \frac{\E \left\{ \left|z \cdot \sum\limits_{i=1}^N
w_{\rm CB-EPA} \cdot a_{i} \right|^2\right\} }{\sigma_{n}^2}
\label{eq23}
\end{eqnarray}
where $w_{\rm CB-EPA}$ is the maximum transmission weight for the
CB-EPA, while the vector of CB weights in the case of CB-EPA is
simply the product of the vector of ones, since for CB-EPA $u_i =
1$, $\forall i$, and $w_{\rm CB-EPA}$. The average SNR
$\bar{\gamma}$ \eqref{eq23} can be then simplified as follows
\begin{eqnarray}
\bar{\gamma} \!\!\!&=&\!\!\!  w_{\rm CB-EPA}^2 \cdot
\frac{\sum\limits_{i=1}^N \E\left\{a_{i}^2 \right\} +
\sum\limits_{i=1}^N \sum\limits_{p = 1,\ p \ne r}^N \E\left\{a_{i}
\right\} \cdot \E\left\{a_p \right\} }{\sigma_{n}^2}
\nonumber \\
\!\!\!&=&\!\!\! w_{\rm CB-EPA}^2 \cdot \frac{ N \cdot (\sigma_{\rm
a}^2 + m_{\rm a}^2) + N \cdot (N-1) \cdot m_{\rm a}^2
}{\sigma_{n}^2} . \label{eq24}
\end{eqnarray}
Solving \eqref{eq24} for $w_{\rm CB-EPA}$, the CB weights can be
found as
\begin{eqnarray}
w_{\rm CB-EPA} = \sqrt{ \frac{ \bar{\gamma} \cdot \sigma_{n}^2}{
N \cdot \sigma_{\rm  a}^2 + N^2 \cdot m_{\rm  a}^2 } }, \quad
\forall r = 1,2, \ldots, N. \label{eq25}
\end{eqnarray}

\section{Factors Affecting lifetime}
CB-PA increases the lifetime of the WSN, however, different
factors affect the efficiency of the power allocation.

\subsection{Initial Energy Distribution}
The energy budget can have different distributions among sensor
nodes in a cluster. We will consider two distributions. One is
uniform distribution where the initial energy can take any value
from zero to the maximum. Gaussian distribution is considered as
well, it represents the case when the initial energy take a
nominal value with some deviation around this value.

\subsection{Single-link and Multi-link CB}
The required bit rate can be achieved using single-link CB where
all collaborative sensor nodes transmit as one cluster.
Alternatively, multi-link CB can be used when we target $K$
BSs/APs located at different directions. In this case, sensor
nodes are grouped into clusters, where each cluster consists of
$N/K$ sensor nodes and targets different destination, thus
establishes independent link. This scheme is called multi-link CB
\cite{Ahmed2010}.

Let ${\cal M}_k$ be a set of nodes targeting BS/AP denoted ${\cal
D}_k$. During the CB step, each collaborative node $c_i, i \in
{\cal M}_k$ transmits the signal
\begin{equation}
t_i = {\it{z}}_k \cdot w_{i} \cdot e^{j\psi_i^k}, \quad i \in
{\cal M}_k
\end{equation}
where $\psi_i^k$ is initial phase in this case, thus
\begin{equation}
\psi_i^k = - \frac{2\pi}{ \lambda} \cdot \delta_i (\varphi_{{\cal
D}_k}).
\end{equation}
Then the received signal at angle $\phi$ from all collaborative
sets of sensor nodes ${\cal M}_k, \; \forall k \in \{0, 1, \dots,
K \}$ can be written as
\begin{eqnarray} \label{ModelMain}
g( \phi ) = \sum\limits_k {\it{z}}_k \cdot \sum\limits_{i \in
{\cal M}_k} w_{i} \cdot a_{ik} \cdot e^{j\psi_i^k} \cdot e^{
\theta_i ( \phi) } + n.
\end{eqnarray}

The received signal at the BS/AP ${\cal D}_l$ can be written as
\begin{eqnarray}
g_{l} &=&  z_{l} \cdot \sum\limits_{i \in {\cal M}_l} w_{i} \cdot
a_{il} + \sum\limits_{k \ne l} {\it{z}}_k \cdot \sum\limits_{i \in
{\cal M}_k} w_{i} \cdot a_{il} \nonumber \\
&\times& e^{j \frac{2 \pi}{\lambda}
\left( \delta_i \left(\varphi_{{\cal D}_k} \right) - \delta_i
\left(\varphi_{{\cal D}_l} \right)\right)} + n. \label{Eq:y1}
\end{eqnarray}
The first term in \eqref{Eq:y1} is the signal received at the
BS/AP ${\cal D}_l$ from the desired set of collaborative nodes
${\cal M}_l$ and the second term represents the interference
caused by other sets of nodes ${\cal M}_k, \forall k \ne l$ with
${\cal M}_k \cap {\cal M}_l = \varnothing$, $l \ne k$. Let the
interference be controlled to a very low level using the methods
of sidelobe control \cite{Ahmed2010}, \cite{MohMe},
\cite{Berbakov2011} and thus can be neglected for simplicity.
Then, \eqref{Eq:y1} reduces to be \eqref{Eq:y1Ch5} and the
difference between single-link and multi-link CB is now only in
the number of collaborative nodes targeting each BS/AP.

The average SNR at the intended BS/AP ${\cal D}_l$ for the
multi-link CB can be found as
\begin{eqnarray} \label{gammaDef_ML}
\bar{\gamma}_{ML} = \E\{ \gamma_{ML} \} = \frac{\E \left\{ \left|z_l
\cdot \sum\limits_{i=1}^{\frac{N}{K}} w_{i} \cdot a_{il} \right|^2
\right\} }{\sigma_{n}^2}, \quad r \in {\cal M}_l.
\end{eqnarray}
Since $K$ source--destination links are used simultaneously, the
average transmission rate of the multi-link CB with node selection
can be expressed as
\begin{eqnarray}
R_{\rm ML} &=& K \log_2 ( 1 + \bar{\gamma}_{ML} ) \qquad {\rm
[bits/sec/Hz]}. \label{Rate1}
\end{eqnarray}

\subsection{Bit Rate}
The logarithmic relation between the SNR and the corresponding bit
rate suggests that reducing the bit rate by few bits will result
in huge reduction in the required SNR, and thus, in reduction of
the consumed energy at each transmission round and increase of the
network lifetime.

\subsection{Number of CB Weight Quantization Levels}
In practice, sensor nodes can transmit with only a finite number
of power levels and thus the CB weights have to be quantized. It
is expected that if the CB weight is allowed only to take few
quantized values, the energy consumption rate will not be
perfectly equalized between different sensor nodes.

\section{Numerical Study}
Numerical simulations are used to illustrate the usefulness of the
proposed CB-PA for extending the lifetime of a cluster of sensor
nodes and study different factors affecting its performance. In
the following examples, we consider a cluster of randomly
distributed sensor nodes over a disk with normalized radius $
 250 \lambda$, where $\lambda$ is the wavelength of the signal
carrier. The BS/AP is located at the direction $\varphi_D = 0^o$ 
and 1 Km away from the sensor nodes. The total number of 
collaborative nodes in the cluster is $N = 100$ and the required 
average SNR $\bar{\gamma}$ at the targeted BS/AP to achieve bit 
rate of 4 [bits/sec/Hz] is $11.76$~dB. Let the noise power to be 
-100 dB and the path-loss ${\rm PL}_0$ at $d_0 = 1$ is 40 dB. 
Assuming path-loss exponent $\alpha = 2$, the required total 
transmit power becomes $P_{\rm Tx} = \bar{\gamma} + {\rm PL}_0 
+ 10 \cdot 2 \cdot \log_{10} (1000) - 100 = 11.76$ dB.  

{{In outdoor WSN applications, i.e., the scenarios in which CB is
needed, sensor nodes are deployed randomly on the ground and the
antenna heights are very low, so that the transmitted signals are
obstructed mainly by the ground \cite{Molina-Garcia-Pardo2005}.
Therefore, we consider the channel characteristics near the ground 
\cite{Sohrabi1999}, \cite{Hugine2006} where, besides the path-loss, 
the large-scale fading has to be considered. In such case the 
fluctuation/shadowing effect of channel coefficient mean appears 
because of hills, forests, buildings, etc. in the signal path.
Typically, the wireless channel attenuation/fluctuation $A_{i}$
for $i$th collaborative node is assumed (accurately) to be a
zero-mean Gaussian distributed random variable (in dB) i.e.,
$A_{i} \sim {\cal N} (0, \sigma^2)$ with $\sigma^2$ being the
variance. The absolute channel gain value has log-normal
distribution, i.e., $a_{i} = 10^{(A_i/10)}$. The variance of 
$A_{i}$ is set to $\sigma^2 = 16$ (in dB). Moreover, sensor nodes 
and any surrounding objects are assumed hereafter to be static, 
which suggests that the channel attenuation varies very slowly 
with time \cite{Molina-Garcia-Pardo2005}, \cite{Zhou2006}.
Therefore, the channel can be considered as constant during each
CB transmission round.}} 

We assume two distributions for the initial energies
assigned to different sensor nodes in a cluster, namely uniform
and Gaussian distributions. {For both distributions, the initial
energies are set between 0 and $E_{\rm{max}} = 1$~J with mean
$m_{e} = E_{\rm{max}}/2 = 0.5$~J.} In the simulations, we assume
that the power allocation is performed every time slot. CB-EPA
\eqref{eq25} are used as the benchmark for comparison. The CB-PA
weights are quantized into $8$ levels. The cluster of sensor nodes
is considered dead when more than $90\%$ of its sensor nodes
deplete energy or the achieved SNR at the targeted BS/AP reduces
by 3~dB below the nominal average value. To show the algorithm
robustness against synchronization errors, we assume that the
initial phase at different sensor nodes is corrupted with uniform
distributed phase error $\Delta \psi \sim {\cal U} [-5^\circ, 5^\circ]$.
All results are averaged over 1000 Monte Carlo runs.

\begin{figure}
\centering
\newlength\figureheight 
\newlength\figurewidth 
\setlength\figureheight{0.4635\textwidth}
\setlength\figurewidth{0.6\textwidth}
\includegraphics{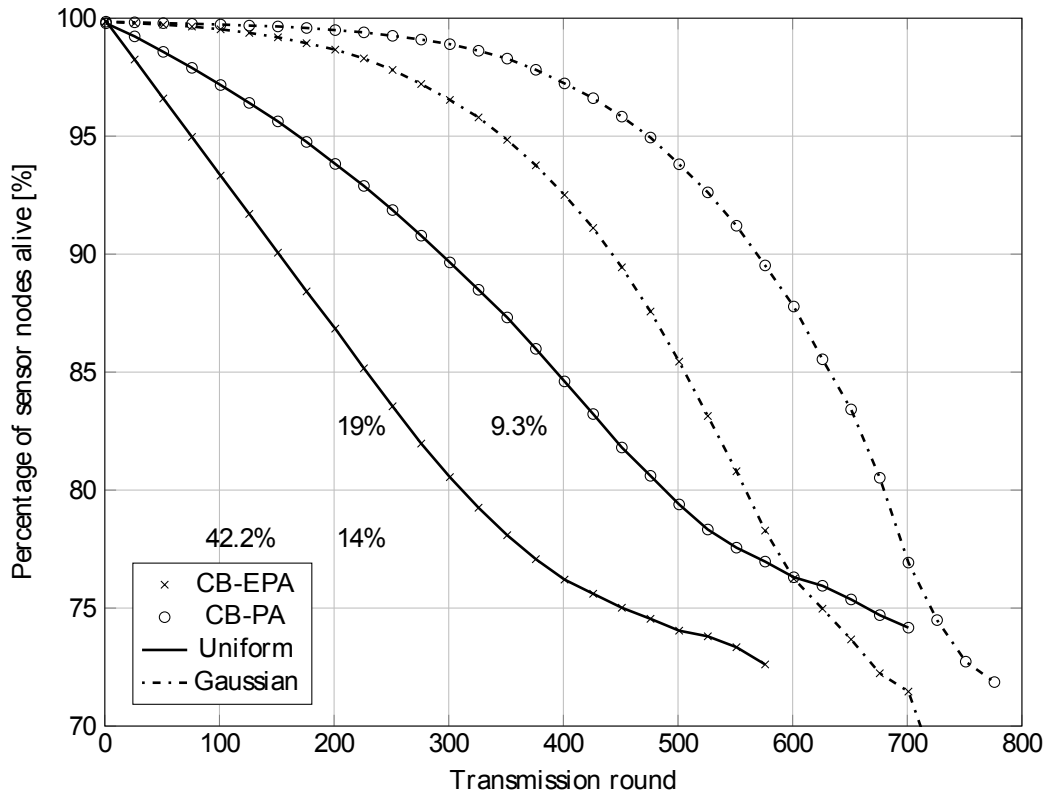}
\caption{Percentage of sensor nodes alive versus time, CB-EPA vs
CB-PA, $\epsilon_{\%}$ is shown for each case.}
\label{Fig2:LifeNodes}
\end{figure}

\begin{figure}
\centering
\setlength\figureheight{0.4635\textwidth}
\setlength\figurewidth{0.6\textwidth}
\includegraphics{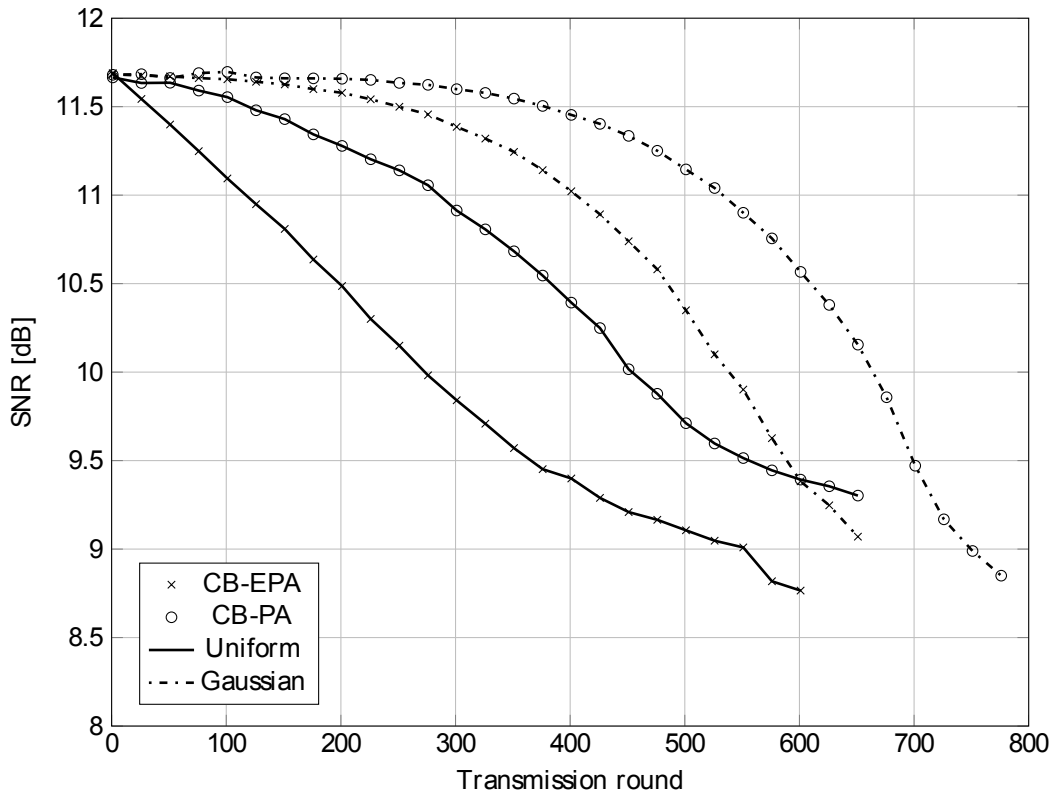}
\caption{Received SNR $\gamma$ at the targeted BS/AP versus time, 
CB-EPA vs CB-PA.} \label{Fig3:SNR}
\end{figure}

{\it Example 1 (Effect of Power Allocation and Initial Energy
Distribution):} In this example, we compare the proposed CB-PA
with the CB-EPA \eqref{eq25}. Fig.~\ref{Fig2:LifeNodes} shows the
percentage of sensor nodes alive versus time for aforementioned CB
methods. The figure shows as well the percentage of total wasted
energy at all sensor nodes $\epsilon_{\%}$. For uniform initial
energy, the percentage of sensor nodes alive decays linearly with
time in the case of CB-EPA because the initial energies at sensor
nodes are not equal. In the case of CB-PA, the percentage of
sensor nodes alive decays at much slower rate. The wasted energy
is reduced to the half of that in the case of CB-PA. This means
that for uniform distribution with CB-EPA, 90\% of sensor nodes
die or are unable to achieve acceptable SNR with about 42\% of
cluster initial energy not used, while 14\% of initial cluster
energy is not used when the cluster dies if the proposed CB-PA is
employed. It clearly demonstrates the advantages of using power
allocation strategy based on which sensor nodes manage to use most
of the cluster energy of sensor nodes assigned for CB. For the
case of Gaussian distributed initial energy, the CB-PA outperforms
the CB-EPA and achieves lower energy depletion rate. Moreover, in
the case of Gaussian distribution, energy depletes slower than in
the case of uniform energy distribution for both CBs tested. This
is due to the fact that for Gaussian distribution, the residual
energy values are concentrated around the mean value $m_{e}$ with
less values closer to 0 and $E_{\rm{max}}$. The wasted energy is
less for the CB-PA as compared to the CB-EPA in the case of
Gaussian energy distribution. In addition, Fig.~\ref{Fig3:SNR}
shows the received SNR at the BS/AP versus time. The SNR
demonstrates similar behavior as the percentage of sensor nodes
alive. Note in Fig.~\ref{Fig2:LifeNodes} that more than 70\% of
the sensor nodes are still alive, however, as Fig.~\ref{Fig3:SNR}
shows, the achieved SNR dropped below the nominal average value
and thus the cluster is considered dead. Note that the energy per 
bit to the noise power spectral density ratio $E_b/N_0$ curves 
can be easily obtained by simply scaling the corresponding SNR 
curves according to the relation 
\begin{eqnarray}
\frac{E_b}{N_0} = \left( \frac{B}{f_b} \right) \gamma
\end{eqnarray}
\noindent where $f_b$ is the bit rate [bits/s] and $B$ 
is the channel bandwidth [Hz].

\begin{figure}
\centering
\setlength\figureheight{0.4635\textwidth}
\setlength\figurewidth{0.6\textwidth}
\includegraphics{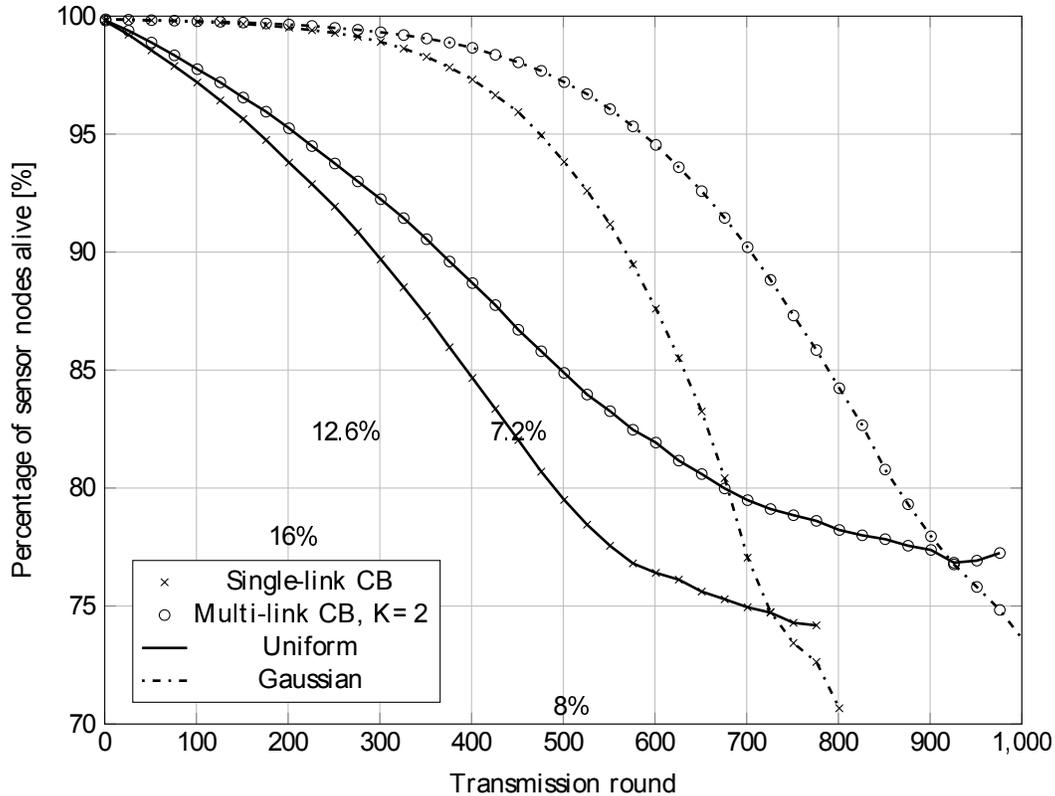}
\caption{Percentage of sensor nodes alive versus time, the
single-link vs multi-link scenario, the wasted energy
$\epsilon_{\%}$ is shown for each case.} \label{Fig4:LifeNodes}
\end{figure}

\begin{figure}
\centering
\setlength\figureheight{0.4635\textwidth}
\setlength\figurewidth{0.6\textwidth}
\includegraphics{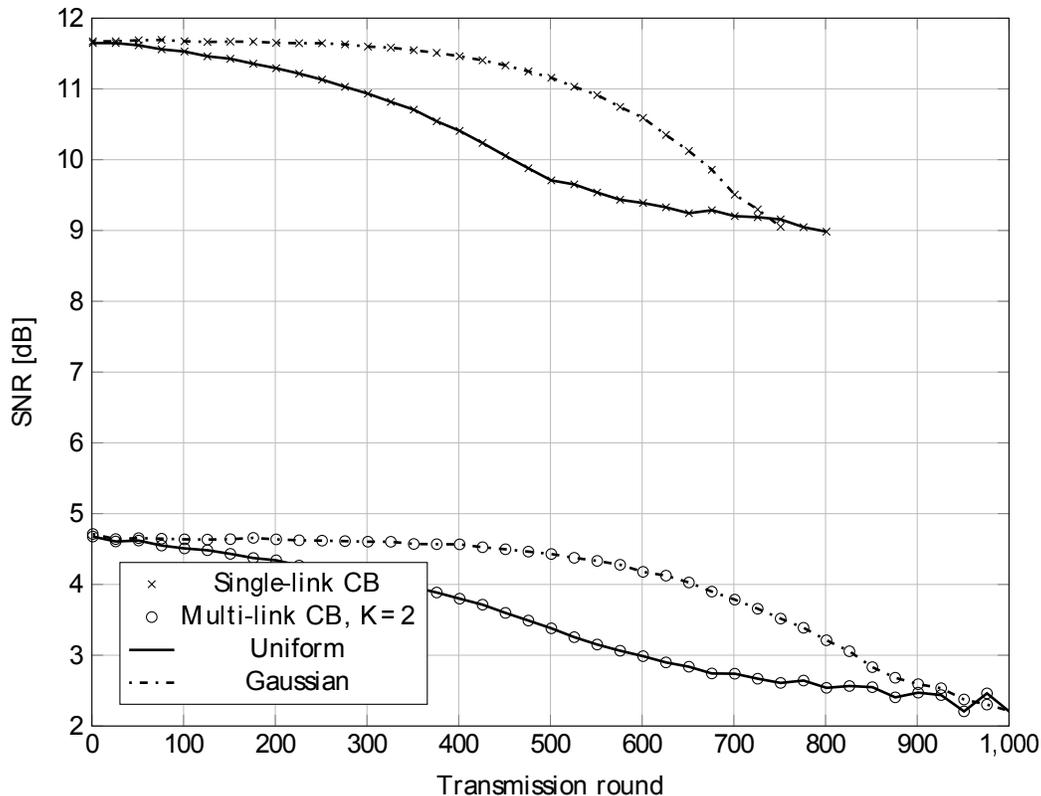}
\caption{Received SNR $\gamma$ at the targeted BS/AP versus time,
the single-link vs multi-link scenario.} \label{Fig5:SNR}
\end{figure}

\begin{figure}
\centering
\setlength\figureheight{0.4635\textwidth}
\setlength\figurewidth{0.6\textwidth}
\includegraphics{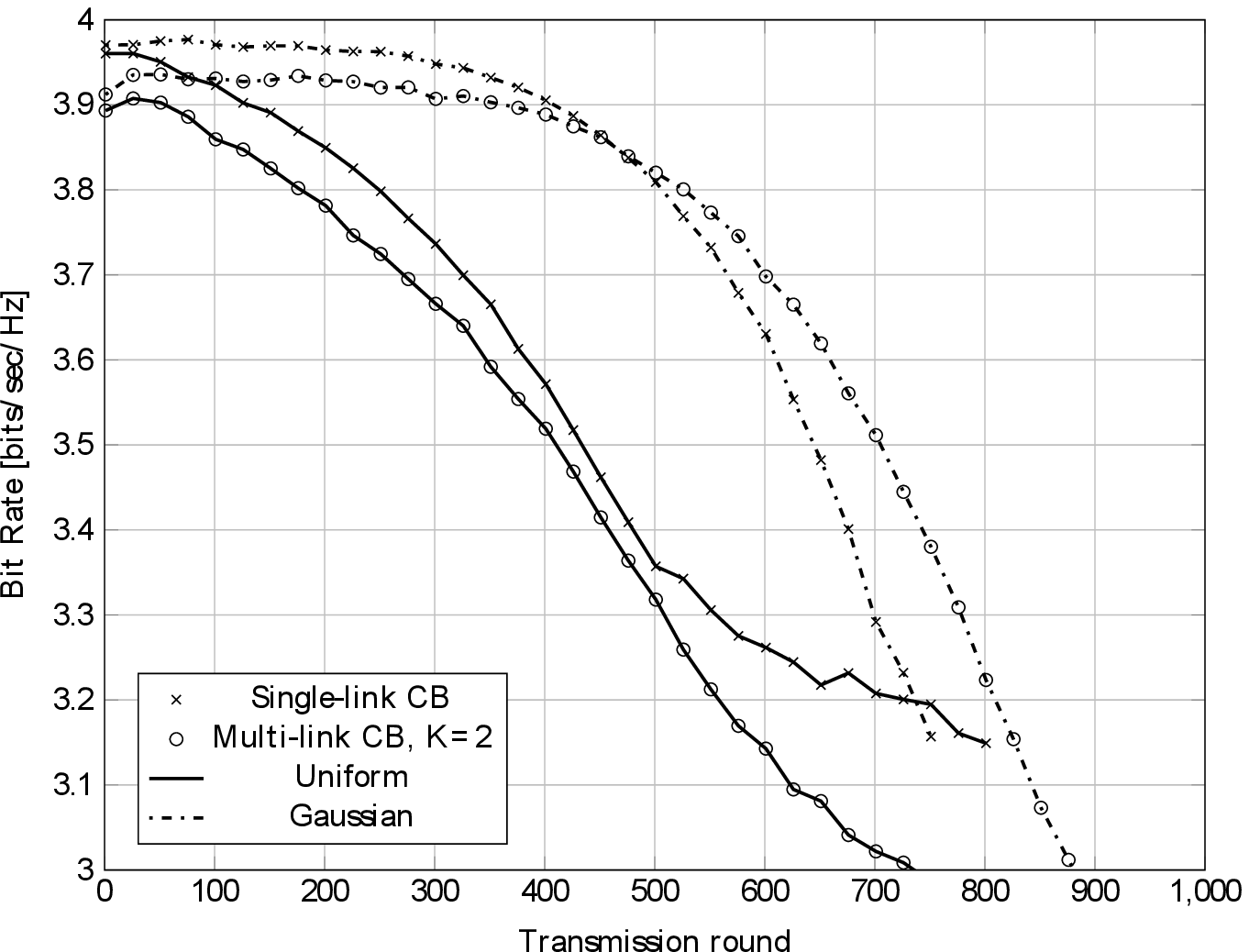}
\caption{Total achieved bit rate versus time, the single-link vs
multi-link scenario.} \label{Fig6:BitRate}
\end{figure}

{\it Example 2 (Effect of Multi-Link CB):} In this example, the
single-link CB is compared to the multi-link CB with 2 clusters
each of them having $N/2=50$ sensor nodes. Power allocation is
used for both aforementioned CB configurations.
Fig.~\ref{Fig4:LifeNodes} shows the percentage of sensor nodes
alive versus time for both aforementioned CB configurations. In
can be seen from the figure that the multi-link CB significantly
extends the lifetime of a cluster of WSN and reduce the wasted
energy. The curve is not smooth close to the end of the lifetime
because fewer clusters survive and the average is taken over them.
Moreover, Fig.~\ref{Fig5:SNR} shows the received SNR at the BS/AP
versus time. The SNR is higher for single-link CB because the
cluster has double the number of sensor nodes. However, the
advantage of using multi-link CB is that we establish independent
links and this increases the achieved bit rate. In addition,
Fig.~\ref{Fig6:BitRate} shows the average bit rate for
single-link and multi-link CB.
 The bit rate is a logarithmic function of the SNR
\begin{eqnarray}
R &=&  \log_2 \left( 1 +  \gamma \right) \qquad {\rm
[bits/sec/Hz]} \label{BitRate}
\end{eqnarray}
and it is expected to decay similar to the SNR. The total bit 
rate of the two links of the multi-link CB is almost equal to 
the bit rate of the single-link CB (each cluster in the multi-link 
CB transmits 2~[bits/sec/Hz]). Moreover, {for the Gaussian case} 
the multi-link CB achieves longer lifetime.

\begin{figure}
\centering
\setlength\figureheight{0.4635\textwidth}
\setlength\figurewidth{0.6\textwidth}
\includegraphics{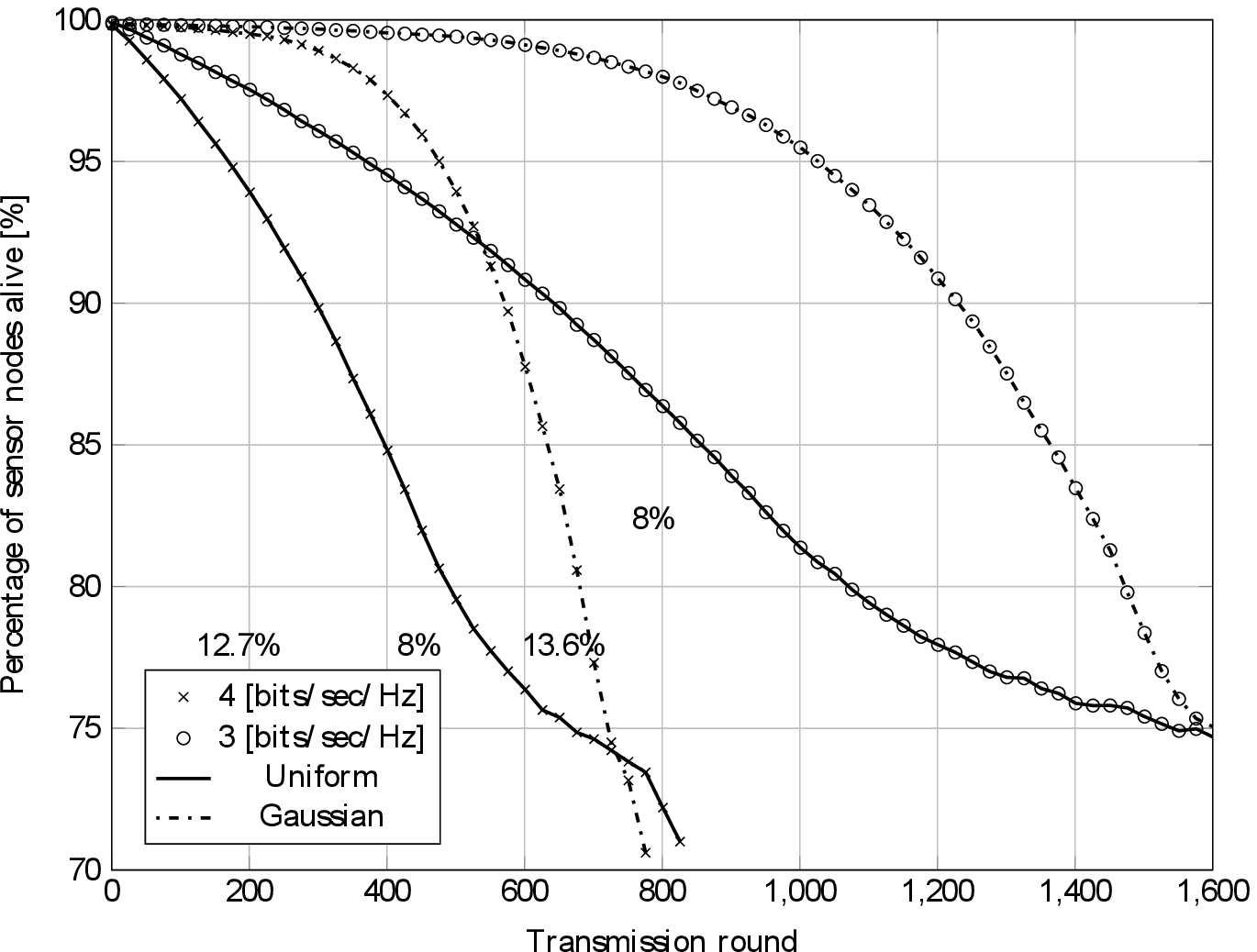}
\caption{Percentage of sensor nodes alive versus time for
different bit rate requirements, the wasted energy $\epsilon_{\%}$
is shown for each case.} \label{Fig7:LifeNodes}
\end{figure}

\begin{figure}
\centering
\setlength\figureheight{0.4635\textwidth}
\setlength\figurewidth{0.6\textwidth}
\includegraphics{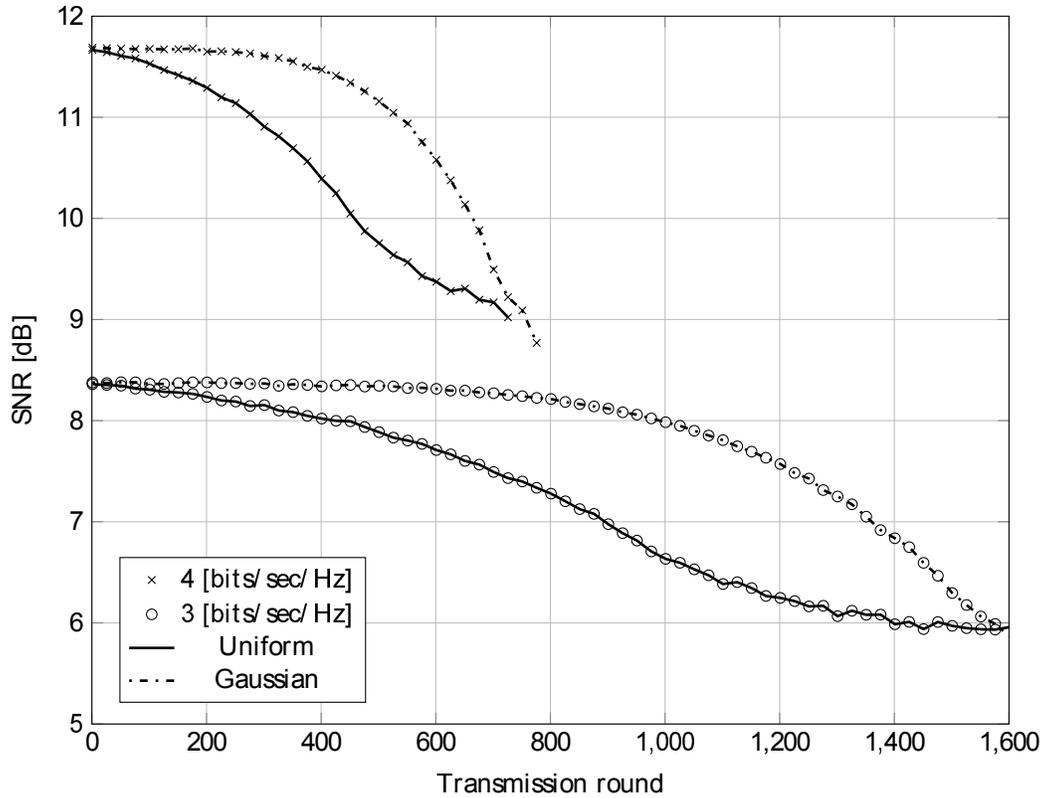}
\caption{Received SNR $\gamma$ at the targeted BS/AP versus time
for different bit rate requirements.} \label{Fig8:SNR}
\end{figure}

\begin{figure}
\centering
\setlength\figureheight{0.4635\textwidth}
\setlength\figurewidth{0.6\textwidth}
\includegraphics{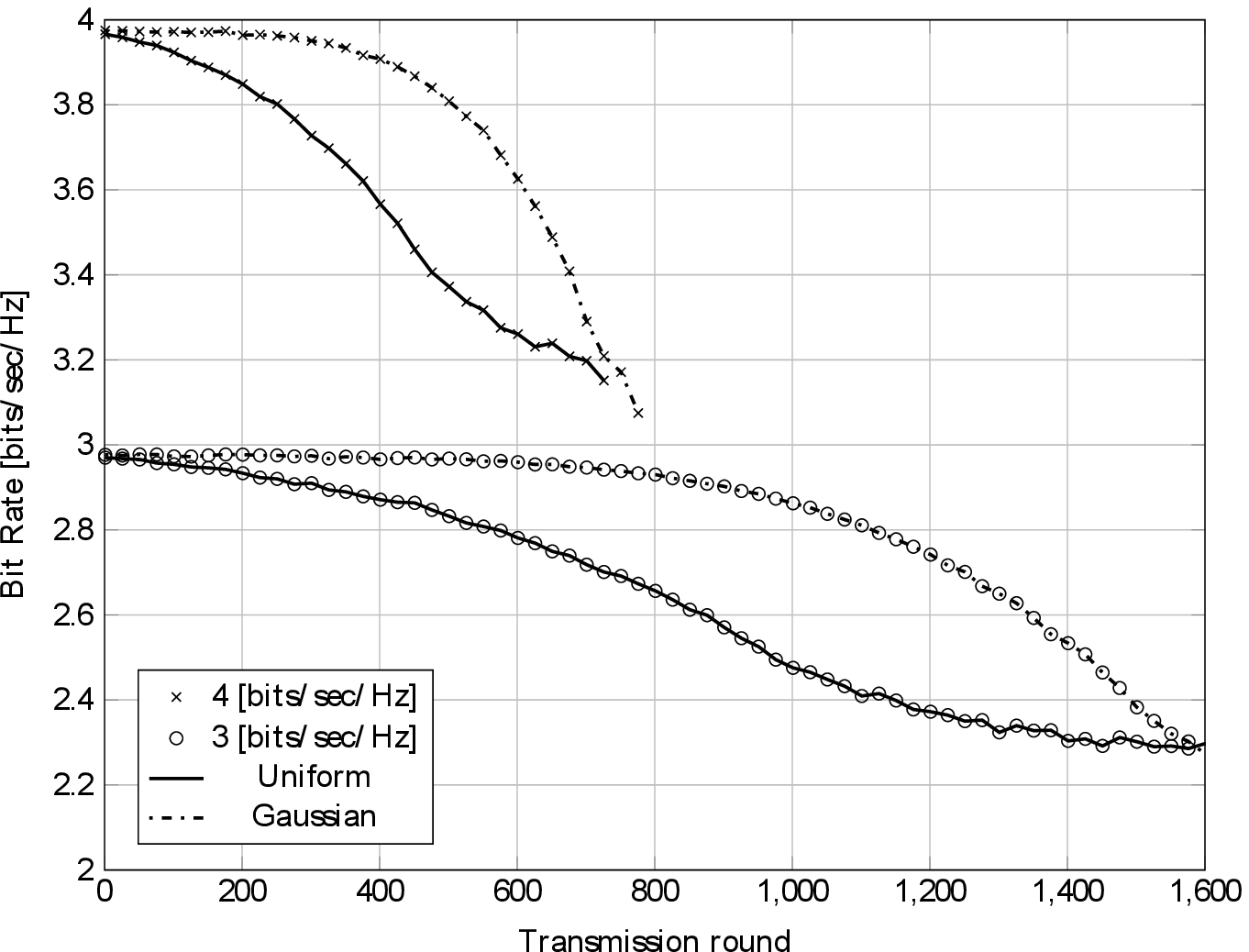}
\caption{Achieved bit rate versus time for different bit rate
requirements.} \label{Fig9:BitRate}
\end{figure}

{\it Example 3 (Effect of Bit Rate):} In this example, we
demonstrate the effect of changing the required bit rate. We
compare transmitting 4 and 3 [bits/sec/Hz], by controlling the
corresponding SNR. Reducing the SNR by 3~dB will result in
reducing the corresponding bit rate by 1 [bit/sec/Hz].
Fig.~\ref{Fig7:LifeNodes} shows the percentage of sensor nodes
alive versus time. In can be seen from the figure that by reducing
the bit rate by 1 [bits/sec/Hz] (and hence the SNR by 3~dB) almost
doubled the cluster lifetime. In addition, Fig.~\ref{Fig8:SNR}
shows the received SNR at the BS/AP versus time and
Fig.~\ref{Fig9:BitRate} shows the corresponding average bit rate.
As expected, the bit rate decays similar to the SNR to the
logarithmic relation.

\begin{figure}[!t]
\centering
\setlength\figureheight{0.4635\textwidth}
\setlength\figurewidth{0.6\textwidth}
\includegraphics{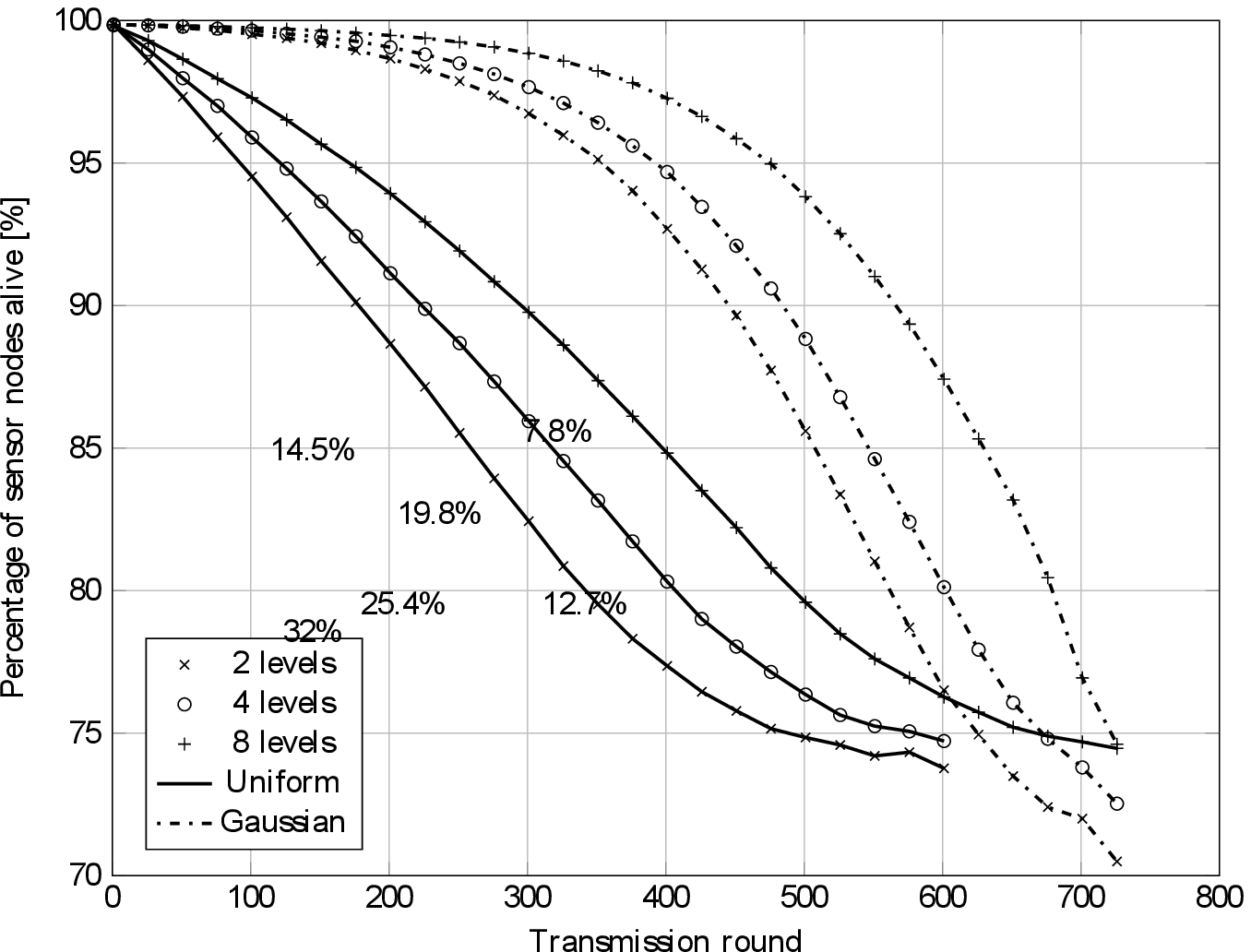}
\caption{Percentage of sensor nodes alive versus time for different 
power quantization levels, the wasted energy $\epsilon_{\%}$ is 
shown for each case.} \label{Fig10:LifeNodes}
\end{figure}

\begin{figure}[!t]
\centering
\setlength\figureheight{0.4635\textwidth}
\setlength\figurewidth{0.6\textwidth}
\includegraphics{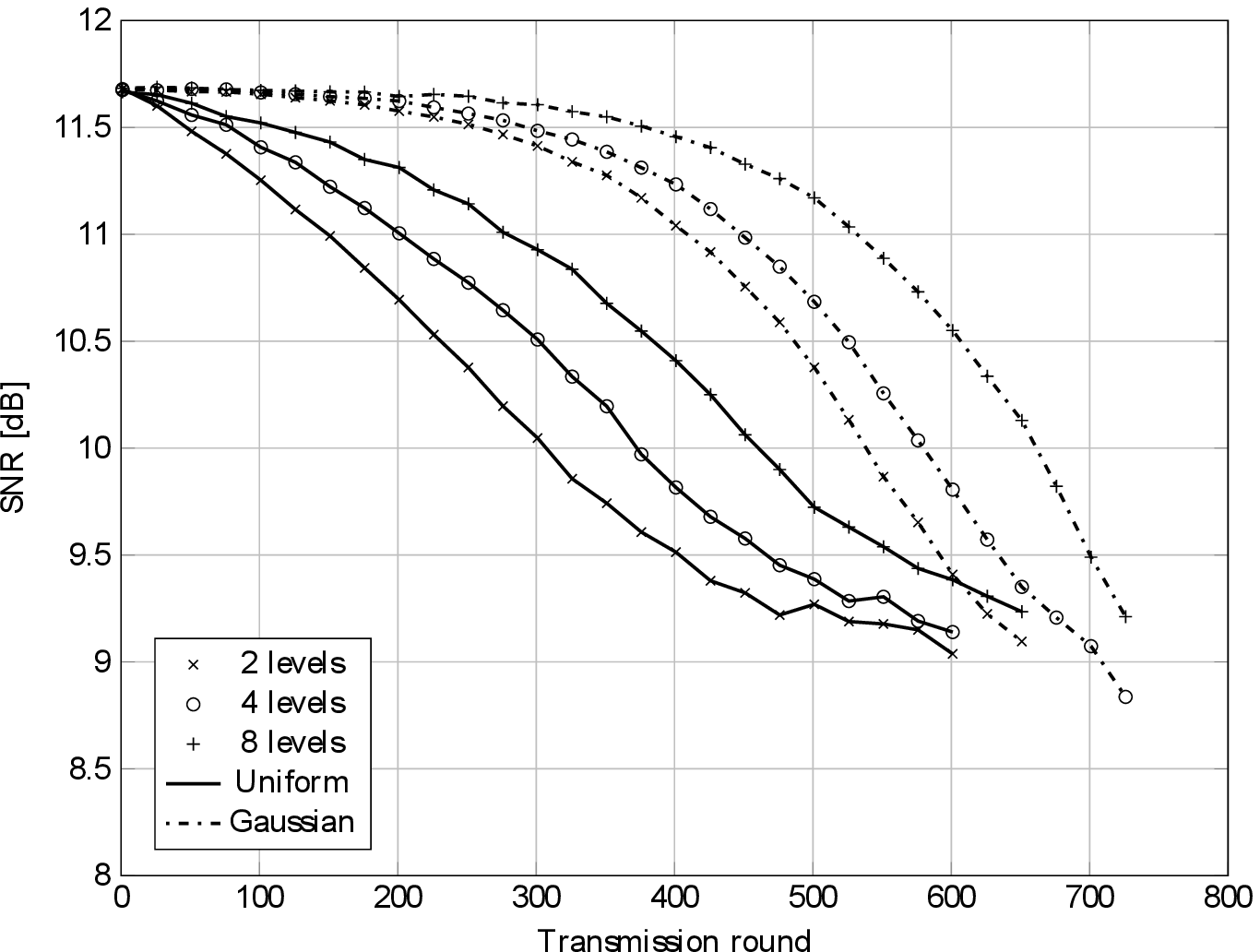}
\caption{Received SNR $\gamma$ at the targeted BS/AP versus time
for different power quantization levels.} \label{Fig11:SNR}
\end{figure}

{\it Example 4 (Effect of the Number of Power Transmission
Levels):} In our final example, we demonstrate the effect of the
number of power transmission levels on the lifetime of a cluster
of sensor nodes. The CB weights are quantized into $2$ (1 bit),
$4$ (2 bits), and $8$ (3 bits) levels which results in $2$, $4$,
and $8$ power levels. Fig.~\ref{Fig10:LifeNodes} shows the
percentage of sensor nodes alive versus time. It is clear from the
figure that having a smoother quantization will extend the
lifetime and reduce the wasted energy. Moreover,
Fig.~\ref{Fig11:SNR} shows the received SNR at the BS/AP versus
time. The SNR shows similar behavior to that of the percentage of
sensor nodes alive in Fig.~\ref{Fig10:LifeNodes}.

\section{Conclusions}
Energy consumption in a cluster of WSN sensor nodes due to CB
transmission has been addressed. In particular, we have presented
an energy model with focus on energy dissipated at the power
amplifier due to CB {{weights}}. Power allocation for CB is
introduced as a PHY layer solution to maximize the lifetime of a
cluster of sensor nodes. The proposed strategy for power allocation
requires only the REI and the statistics of both the channel gains
and the energy available at the cluster and, thus, can be
classified as semi-distributed. In addition to the fact that the
proposed strategy is simple, simulation results illustrate very
significant improvements of the cluster lifetime due to the CB
power allocation. The proposed algorithm does not necessarily
guarantee globally optimum CB power allocation, which can be
achieved by solving the problem centrally. However, it achieves
excellent results with low implementation complexity which is
suitable for WSNs. Different factors affecting the lifetime of a
cluster of WSN are also discussed and analyzed by simulations. 
Moreover, the multi-link CB increases the
lifetime compared to the single-link CB. In addition, reducing the
SNR and hence the corresponding bit rate results in increasing the
lifetime and thus the total transmitted bit rate during the
lifetime. Having more control on the transmission power values,
that is, introducing more quantization levels for CB weights
(power levels), improves the lifetime as well.

\end{document}